\documentclass[aps,twocolumn,superscriptaddress,prl]{revtex4}
\usepackage{amsmath}
\usepackage{graphicx}
\begin{document}

\newcommand{\Tr}{\text{Tr}}

\title{Neutron wave packet tomography}

\author{G. Badurek}
\affiliation{Atominstitut der \"{O}sterreichischen
Universit\"{a}ten, Stadionallee 2, A-1020 Wien, Austria}
\author{P. Facchi}
\affiliation{Dipartimento di Fisica, Universit\`a di Bari, I-70126
Bari, Italy}
\affiliation{Istituto Nazionale di Fisica Nucleare, Sezione di Bari, I-70126
Bari, Italy}
\author{Y. Hasegawa}
\affiliation{Atominstitut der \"{O}sterreichischen
Universit\"{a}ten, Stadionallee 2, A-1020 Wien, Austria}
\author{Z. Hradil}
\affiliation{Department of Optics, Palack\'y University, 17.
listopadu 50, 772 00 Olomouc, Czech Republic}
\author{S. Pascazio}
\affiliation{Dipartimento di Fisica, Universit\`a di Bari, I-70126
Bari, Italy}
\affiliation{Istituto Nazionale di Fisica Nucleare, Sezione di Bari, I-70126
Bari, Italy}
\author{H. Rauch}
\affiliation{Atominstitut der \"{O}sterreichischen
Universit\"{a}ten, Stadionallee 2, A-1020 Wien, Austria}
\author{J. \v{R}eh\'{a}\v{c}ek}
\affiliation{Department of Optics, Palack\'y University, 17.
listopadu 50, 772 00 Olomouc, Czech Republic}
\author{T. Yoneda}
\affiliation{Dipartimento di Fisica, Universit\`a di Bari, I-70126
Bari, Italy} \affiliation{School of Health Sciences, Kumamoto
University, 862-0976 Kumamoto, Japan}


\begin{abstract}
A tomographic technique is introduced in order to determine the
quantum state of the center of mass motion of neutrons. An
experiment is proposed and numerically analyzed.
\end{abstract}


\maketitle

In experimental physics one often faces the following question:
``Given the outcomes of a particular set of measurements, which
quantum state do they imply?'' Such {\em inverse problems} may
arise for instance when setting up and calibrating laboratory
sources of quantum states, or in the analysis of decoherence and
other deteriorating effects of the environment, or in some special
tasks in quantum information processing such as eavesdropping on a
quantum channel in quantum cryptography.

The determination of the quantum state represents a highly
nontrivial problem, whose history can be traced back to the early
days of quantum mechanics, namely to the Pauli problem
\cite{St,St2}; the experimental validation had to wait until quantum
optics opened a new era. The theoretical predictions of Vogel and
Risken \cite{VR} were closely followed by the experimental
realization of the suggested algorithm by Smithey \emph{et al.}
\cite{SBRF93}. Since then many improvements and new
techniques have been proposed: an up-to-date overview can be found
in Ref.\ \cite{lnp}. Recent progress in instrumentation has made
it possible to apply these techniques to a variety of different
quantum systems such as fields in optical cavities, polarization
and external degrees of freedom of photons, or motional states of
atoms.

In this Letter we propose an experiment for determining the
quantum states of the center-of-mass motion of neutrons. In
accordance with quantum theory, these massive particles can be
associated with a wave function describing their motional state.
Neutrons are suitable objects for many quantum mechanical
experiments due to their interaction with all four basic forces,
the ease of detecting them with almost $100\%$ efficiency, and
their small coupling to the environment \cite{rauch-book}. In
marked contrast with light, neutron vacuum field and thermal
background can usually be ignored. This makes it possible, for
instance, to prepare superpositions of macroscopically separated
quantum states---the so-called Schr\"odinger cat states---that
would be extremely difficult to realize with other quantum systems
because of their fragility with respect to decoherence. In all
experiments performed so far, the existence of the Schr\"odinger
cat states of neutrons has been indirectly demonstrated via
interferometric effects, but the full evidence for the
nonclassicality of these states, including the presence of the
negative values of the reconstructed Wigner function, is still
missing.

In the following, we will first briefly review the present neutron
interferometric techniques and the means of creating highly
nonclassical motional states of neutrons. In the second part of
the Letter, an experiment will be proposed for the complete
reconstruction of these quantum states.


\underline{Neutron tomography} - The set of measurements that can
be done on neutrons to determine their quantum state is severely
limited by the very low time resolution of the available
detectors. In quantum optics, this obstacle can be overcome by
mixing the weak input field with a strong local oscillator. By
changing the phase $\phi$ of the oscillator one can measure the
spectral decompositions of all quadratures,
\begin{equation}\label{oscil}
\hat X_\phi=\hat x \cos\phi+\hat p \sin\phi,
\end{equation}
$\hat x$ and $\hat p$ being the canonically   conjugated operators
of position and momentum. Of course, no such local oscillators
exist for neutrons. However, notice that massive particles
experience a transformation of the type \eqref{oscil} in the
course of free evolution: $x(t)=x +(p/m)t$, where $m$ is the mass.
Thus free evolution of the wave packet followed by a position
sensitive measurement yields information about a subset of
quadratures $X_\phi$, $\phi\in[0,\pi/2]$. Free evolution was
utilized e.g.\ for the reconstruction of transversal motional
states of helium atoms in \cite{konst}. Here we are interested in
the {\it longitudinal} degrees of freedom. Since neutron detectors
have very bad time resolution, free evolution {\it alone} cannot
be used to generate a tomographically complete set of
measurements.

\begin{figure}
\centerline{\includegraphics[width=0.8\columnwidth]{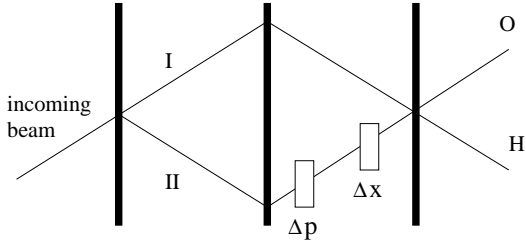}}
\caption{Scheme of a perfect crystal neutron interferometer. The
incoming beam is split at the first crystal plate, reflected at
the middle plate and recombined again at the third plate. The
detector is placed in beam O where the visibility is higher due to
the same number of reflections/transmissions. In addition to a
position shift $\Delta x$ routinely used in neutron experiments, a
momentum kick $\Delta p$ has been added in path II in order to
make the interferometric measurement tomographically complete; see
text.\label{fig:pecnoxp}}
\end{figure}

Feasible measurements on thermal neutrons consist of measurements
of the contrast and phase of interference fringes in an
interferometric setup, see Fig.~\ref{fig:pecnoxp} (without
momentum kick), and also spectral analysis of the neutron beam
using an adjustable Bragg-reflecting crystal plate together with a
position sensitive detector. This set of observables is
\textit{not} tomographically complete because the measurable
(complex) contrast of the interference pattern \cite{rauch-book}
($\hbar=1$),
\begin{equation}
\Gamma(\Delta x)=\langle \psi|e^{i\Delta x \hat p}|\psi\rangle=
\int |a(p)|^2 e^{i \Delta x  p} dp,
\end{equation}
is not sensitive to the phase of $a(p)=\langle p |\psi\rangle$,
and no information about quadratures other than $p$ is available.

Obviously, the situation would be different if one could shift
both the position (phase) \textit{and} the momentum of the
incoming wave packet inside the interferometer. Such a thought
experiment is shown in Fig.~\ref{fig:pecnoxp}. In that case, the
Wigner function describing the ensemble of measured neutrons would
be related to the measured contrast
\begin{equation}\label{gencontrast}
\begin{split}
\Gamma(\Delta p,\Delta x)&= \mathrm{tr}\left\{\rho e^{i\Delta p
\hat x}e^{i\Delta x \hat p}\right\}\\
&=\int e^{i (\Delta p) x}\langle x|\rho| x+\Delta x\rangle dx
\end{split}
\end{equation}
by a simple integral transformation,
\begin{equation}\label{inversion}
W(x,p)=\int\!\!\!\int e^{-i\frac{uv}{2}+iux+ivp}\Gamma(-u,v)dudv,
\end{equation}
where $u=\Delta p$, $v=\Delta x$ and $\rho$ is the state to be
reconstructed.

Although this thought experiment looks simple, its experimental
realization, according to Fig.~\ref{fig:pecnoxp}, would be rather
difficult. Large momentum kicks acquired by the neutron in the
lower arm would change its de Broglie wavelength and spoil the
Bragg reflection at the last crystal plate. Therefore we will now
propose a modified scheme that can substitute the interferometric
setup of Fig.~\ref{fig:pecnoxp}.

\begin{figure}
\centerline{
\includegraphics[width=\columnwidth]{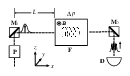}
} \caption{Setup for the tomography of motional states of
neutrons. M$_1$, M$_2$ -- magnetic mirrors; L -- region of free
propagation; B -- static magnetic field controlling the momentum
kick $\Delta p$; F -- box containing a static magnetic field
(aligned along $+y$) and an RF coil; D -- detector; the state is
``prepared" by P, which can be a chopper or a monochromator. The
arrows denote the polarization of neutrons after reflections from
magnetic mirrors.
\label{fig:realistic}}
\end{figure}

\underline{Setup} - In the new scheme shown in
Fig.~\ref{fig:realistic}, the incoming neutrons polarized in the
$+z$ direction, $|\Psi\rangle=|\psi\rangle|z_+\rangle$, where
$\psi$ denotes the spatial degrees of freedom, first propagate
freely through a distance $L$, undergoing a unitary operation
$U_1=\exp[-i \hat p^2 L/(2p_0)]$. In the following, we will assume
that the input wave packets are quasi-monochromatic, $\sigma_p/p_0
\simeq$ a few percent, with spread $\sigma_p$ and
central momentum $p_0$. This condition guarantees that the action
of the RF coil is practically eqivalent to a momentum ``kick." The
generalization to more general states will be presented elsewhere.
After the region of free propagation, the neutrons are let through
an RF coil placed in a static magnetic field polarized along the
$-y$ direction, see Fig.~\ref{fig:realistic}. As a result of the
interaction between the neutron and the coil, the $y_+$ ($y_-$)
component of the input state will be decelerated (accelerated).
Assuming that the region of interaction is short, so that the
dispersion of the wave packet of the neutron can be neglected in
the coil, in the quasi-monochromatic approximation, the net
momentum transfer can be described by the effective unitary
operator,
\begin{equation}\label{opU2}
U_2=e^{-i\Delta p \hat x/2}|y_-\rangle\langle y_+|+e^{i\Delta p
\hat x/2}|y_+\rangle\langle y_-|,
\end{equation}
where
\begin{equation}\label{deltap}
\Delta p=\frac{2\mu B m}{p_0}.
\end{equation}
Prior to detection, the particles are polarized along the $+z$
direction again, so as to erase the which-way information stored
in the polarization degree of freedom. The probability of a
neutron being detected is given by the norm of the transmitted
component,
\begin{equation}\label{PP}
P=\Tr\{\Pi(\Delta p,\Delta x)\rho\} ,
\end{equation}
where $\rho$ refers only to the spatial degrees of freedom and
\begin{eqnarray}
\Pi(\Delta p,\Delta x) &=& \langle z_+|U_1^\dag
U_2^\dag|z_+\rangle
\langle z_+|U_2 U_1|z_+\rangle  \label{povm_setup0} \\
&=& (1+ e^{i \frac{L}{2 p_0}\hat p^2} e^{i\Delta p
\hat{x}}e^{-i
\frac{L}{2p_0}\hat p^2})/4+ \mbox{h.c.} \nonumber \\
&=& (1+e^{i \Delta p(\hat{x}+ L \hat{p}/p_0)})/4+\mbox{h.c.}
\label{povm_setup}\\
&=& (1+e^{i \Delta x\hat{p}} e^{i \Delta p \hat{x}}e^{-i\Delta
x\Delta p/2})/4+\mbox{h.c.},\qquad
\label{povm_setup1}
\end{eqnarray}
where we denoted
\begin{equation}\label{deltax}
\Delta x=\frac{\Delta p L}{p_0}=\frac{2\mu B m L}{p_0^2}.
\end{equation}

\underline{Direct inversion} - The detection probability
(\ref{PP}) reads
\begin{equation}\label{probab-det}
P(\Delta p,\Delta x)=\frac{1}{2}+\frac{1}{2}
\mathrm{Re}\big\{\Gamma(\Delta p,\Delta x)e^{2i\Delta x\Delta
p}\big\}.
\end{equation}
Since the beam is quasi-monochromatic one has for $\delta x=\pi/2
p_0$,
\begin{equation}\label{imaginary}
\Gamma(\Delta p,\Delta x+\delta x)\simeq\Gamma(\Delta p,\Delta
x)e^{i \pi/2}
\end{equation}
from which the imaginary part of the complex degree of coherence
$\Gamma(\Delta p,\Delta x)$ can be obtained.

Summarizing, the tomography of a neutron state consists in the
following four steps:

\noindent (i) A set of pairs of independent variables
$\{B_j,L_j\}$ is chosen covering a certain range $B\in [0,
B_\text{max}]$ and $L\in [0,L_\text{max}]$.

\noindent (ii) For each pair $B_j, L_j$ the shifts $\Delta p_j$ in
 (\ref{deltap}) and $\Delta x_j$ in (\ref{deltax}) are calculated,
and the corresponding intensities $P(\Delta p_j,\Delta x_j)$ are
measured with and without an auxiliary shift $\delta x=\pi/2$.

\noindent (iii) The complex degree of coherence $\Gamma(\Delta
p_j, \Delta x_j)$ is calculated from the two intensities using
Eqs.~\eqref{probab-det}-(\ref{imaginary}).

\noindent (iv) Finally, the Wigner function of the input neutrons
is calculated with the help of inversion formula
\eqref{inversion}, where the integrals are approximated by
sums over $\Delta x_j$ and $\Delta p_j$.

According to (\ref{inversion}), the contrast $\Gamma(\Delta p,
\Delta x)$ is essentially the Fourier transform of the Wigner
function $W(x,p)$. Therefore the largest values of $\Delta p$ and
$\Delta x$ are related to the smallest resolved details in $x$ and
$p$ respectively. Namely (reinserting $\hbar$),
\begin{equation}\label{eq:resolution}
\Delta p_{\mathrm{MAX}}=\hbar/\delta x_{\mathrm{min}}, \quad
\Delta x_{\mathrm{MAX}}=\hbar/\delta p_{\mathrm{min}},
\end{equation}
where $\delta x_{\mathrm{min}}$ and $\delta p_{\mathrm{min}}$
denote the $x$ and $p$ resolutions. By (\ref{deltap}) and
(\ref{deltax}) one gets
\begin{equation}\label{eq:resolution1}
\delta x_{\mathrm{min}}= \frac{\hbar}{2 \mu m}
\frac{p_0}{B_{\mathrm{MAX}}}, \qquad \delta
p_{\mathrm{min}}=\frac{p_0}{L}\delta x_{\mathrm{min}} .
\end{equation}
For a neutron of wavelength $\lambda_0=0.37$nm  \cite{BadurekOC},
assuming the reasonable values $L_{\mathrm{MAX}}=1$m and
$B_{\mathrm{MAX}}=0.1$T one gets $\delta x_{\mathrm{min}}=60
\mu$m and $\delta p_{\mathrm{min}}= \hbar \times
10^{6}$m$^{-1}$.

\underline{Radon inversion} - It is interesting to give an
alternative interpretation of the proposed measurement in
Fig.~\ref{fig:realistic}. Notice, that the POVM elements in
Eq.~\eqref{povm_setup} can also be restated in terms of quadrature
operators,
\begin{equation}\label{quadratures}
\Pi(\Delta p,\Delta x)=(1/4)\big(1+e^{i\omega \hat
X_\theta}\big)+\mbox{h.c.},
\end{equation}
where (in fixed units)
\begin{equation}\label{par_def}
\hat X_\theta=\cos\theta\hat x+\sin\theta \hat p,\quad
\tan\theta=\frac{\Delta x}{\Delta p}=\frac{L}{p_0},
\end{equation}
and $\omega=\sqrt{\Delta x^2+\Delta p^2}$. Thus, for a fixed
$\theta$, the data contain information about the characteristic
function of the quadrature $\hat X_\theta$,
\begin{eqnarray}\label{character}
P(\Delta p,\Delta x)&=&1/2+\text{Re}\{ C_{X_\theta}(\omega)\}/2, \\
\label{char_def}
\langle C_{X_\theta}(\omega)\rangle&=&\int P_{X_\theta}(x)e^{i\omega
x}dx.
\end{eqnarray}
By changing $L$ one changes the quadrature measured, while
$\omega$, which depends on both $L$ and $B$, determines the
observed spatial frequency of the probability distribution of this
quadrature. The observed quadratures range from $\hat x$ (for
$L=0$) to $\hat p$ (for $L\rightarrow \infty$). From the
measurement of $C_{X_\theta}(\omega)$, the ``shadows''
$P_{X_\theta}(x)$ of the Wigner function can be obtained by the
Fourier transformation, which in turn yield the Wigner function by
an inverse Radon transformation. This is an alternative way of
reconstruction the Wigner function from the measured data in the
setup Fig.~\ref{fig:realistic}.

\underline{Statistical inversion} - The procedures outlined above,
based on the direct inversion formula \eqref{inversion}, have
several drawbacks: (i) Realistic data are always noisy. In that
case, formula \eqref{inversion} can yield unphysical results, such
as the Wigner representation of a non-positive definite operator.
(ii) The Wigner function in Eq.\ \eqref{inversion} depends on the
measured data indirectly, through the complex degree of coherence
$\Gamma$, which itself has to be estimated with the help of an
auxiliary position shifter. This intermediate step is, certainly,
not necessary as all available information about the Wigner
function of the incoming neutrons is contained in the raw data
measured without any auxiliary position shift. In order to avoid
these problems, we propose to use the maximum-likelihood quantum
state reconstruction \cite{Zdenek1,Jarda1,lnp}. The main
advantages of this method compared to the above direct inversion
are: (i) Asymptotically, for large data samples it provides the
best performance available. (ii) Any prior information about the
measured neutrons and the known statistics of the experiment can
be used to increase the accuracy of the reconstruction. (iii) The
existing physical constraints can be easily incorporated into the
reconstruction. Most notably, this technique guarantees the
positivity of the reconstructed density operator. (iv) It can be
applied directly to raw counted data.

Assuming that the statistics of the experiment is Poissonian, the
maximum-likelihood reconstruction amounts to minimizing the
Kullback-Leibler distance (relative entropy) between the measured
data $f(\Delta x,\Delta p)$ and the renormalized theoretical
probabilities $p(\Delta x,\Delta p)/\sum p$ of
Eq.~\eqref{probab-det}. As has been shown in
\cite{Zdenek1,Jarda1}, the maximum-likely density matrix can be
obtained as a fixed point of the iterations of a nonlinear
operator map.

\begin{figure}
\centerline{
\includegraphics[width=\columnwidth]{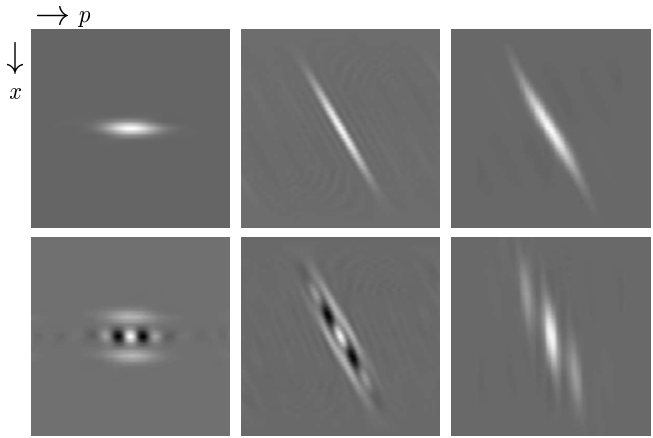}
} \caption{Reconstructed Wigner functions of Gaussian states
(upper row) and superpositions of Gaussian states (``cats;" lower
row) from the simulated data. A $50\times 50$ matrix of $\Delta x$
and $\Delta p$ shifts was used for the maximum-likelihood
inversion. Left column: reconstructed original states; middle
column: reconstructed time-evolved states with a resolution
$\delta x_{\rm min}/l_{\rm coh}= 1/2$; right column: reconstructed
time-evolved states with reduced resolution, $\delta x_{\rm
min}/l_{\rm coh}= 10$. Grey is the zero level, white (black)
represents positive (negative) values.
\label{fig:iter}}
\end{figure}

As follows from the parameter estimates given after
Eq.~\eqref{eq:resolution1}, the proposed tomography scheme using
thermal neutrons will likely have sufficient resolution in
momentum. On the other hand, even for well monochromatized thermal
beams, the resolution in position is expected to be worse
(possibly even by several orders of magnitude) than typical
coherence lengths. The simulations in Figure~\ref{fig:iter}
illustrate the effect of the restricted range of  $\Delta p$ on
the reconstruction.
Consider first the reconstruction of a minimum uncertainty
Gaussian wave packet in its moving frame, parameterized by its
coherence length $l_{\rm coh}$, $|\Psi_G\rangle\propto \int
\exp(-k^2 l_{\rm coh}^2)|k\rangle dk$. (The choice of a
minimum uncertainty state is only for illustrative purposes.)
Provided the apparatus has a sufficient spatial resolution,
$\delta x_{\rm min}<l_{\rm coh}$, a faithful reconstruction is
readily obtained, see the upper left panel. More realistic
measurement with $\delta x_{\rm min} > l_{\rm coh}$ would
obviously yield a Wigner function smoothed out along the $x$ axis.
However, the states measured in a real experiment are not going to
be minimum uncertainty states. The experimenter will rather deal
with time evolved states $|\Psi(T)\rangle\propto
 \int \exp(i k^2 T/2m-k^2 l_{\rm coh}^2)|k\rangle dk$ that are
strongly affected by dispersion. As a consequence, the wave packet
spread very soon becomes larger than the resolution limit, $\delta
x_T\sim T/(m l_{\rm coh})
\gg\delta x_{\rm min}$, and a good reconstruction can be
achieved with a realistic apparatus. Compare the upper middle and
right panels, showing reconstructions with a sufficient resolution
$\delta x_{\rm min}= l_{\rm coh}/2$ and a reduced (but more
realistic) resolution $\delta x_{\rm min}= 10 l_{\rm coh}$.

The imaging of non-classical states is a much more delicate
problem. Let us consider the superpositions of spatially separated
Gaussian states (Schr\"odinger cats), $|\Psi_{\rm cat}\rangle
\propto [1+\exp(i\hat p \Delta)]|\Psi_G\rangle$.
Such states can be prepared e.g.\ by means of a double loop
perfect crystal interferometer \cite{double}. As has been shown
\cite{BadurekOC}, the preparation of thermal neutron cat states is
possible, with separations exceeding the corresponding coherence
lengths of the individual components $\Delta\gg l_{\rm coh}$.
Provided the apparatus has sufficient resolution, the
nonclassicality of this state is manifested by the negative
regions of the reconstructed Wigner function, see the lower left
panel of Fig.~\ref{fig:iter}. Taking dispersion into account,
the ordering of the relevant parameters is $l_{\rm coh}< \Delta <
\delta x_T$, and one can easily obtain $\delta x_{\rm min}<
\delta x_T$. As the simulations show, a realistic measurement
whose position resolution is much worse than the coherence lengths
of the individual cat state components tends to wipe out the
negative regions of the reconstructed Wigner function; compare the
lower middle and right panels of Fig.~\ref{fig:iter}. On the other
hand, the main features of such exotic states, such as their
non-Gaussian character and also the global spatial properties of
which little is known today, should still be accessible to a
realistic wave packet tomography. To resolve more subtle quantum
interference effects of the order of the coherence length, more
refined experimental techniques may however be needed. An idea
could be to replace thermal neutrons by ultracold neutrons, for
which much larger momentum shifts $\Delta p$ (and thus much
smaller $\delta x_{\rm min}$, possibly even smaller than $\Delta$)
can be obtained.

In conclusion, we have proposed and analyzed an experimental
scheme for determining the motional states of neutrons. With the
help of a magnetic field and free propagation, this apparatus
realizes quadrature measurements on neutrons by measuring overlaps
of the two transformed components of the initial state. This is an
analog of the quantum optical homodyne detection in neutron
optics, achieved without the use of a strong coherent source of
neutrons.


\acknowledgments
This work was supported by the Bilateral Research Program between
Italy and the Czech Republic PH1 on ``Decoherence and Quantum
Measurements" and by the Research Project MSM 6198959213 of the
Czech Ministry of Education.

\vspace{-0.3cm}

\bibliography{reconstruct}

\end{document}